\documentclass[
 reprint,
 amsmath,amssymb,
 aps,showpacs,prl,letterpaper,braket]{revtex4-1}
\usepackage{braket}
\usepackage{graphicx}
\usepackage{dcolumn}
\usepackage{bm}
\usepackage{epstopdf}

\begin{document}

\title{Spin-Orbit Coupling and Spin Textures in Optical Superlattices}

\author{Junru Li}
\thanks{These two authors contributed equally.}
\affiliation{Research Laboratory of Electronics, MIT-Harvard Center for Ultracold Atoms, Department of Physics,
Massachusetts Institute of Technology, Cambridge, Massachusetts 02139, USA}

\author{Wujie Huang}
\thanks{These two authors contributed equally.}
\affiliation{Research Laboratory of Electronics, MIT-Harvard Center for Ultracold Atoms, Department of Physics,
Massachusetts Institute of Technology, Cambridge, Massachusetts 02139, USA}

\author{Boris Shteynas}
\affiliation{Research Laboratory of Electronics, MIT-Harvard Center for Ultracold Atoms, Department of Physics,
Massachusetts Institute of Technology, Cambridge, Massachusetts 02139, USA}

\author{Sean Burchesky}
\affiliation{Research Laboratory of Electronics, MIT-Harvard Center for Ultracold Atoms, Department of Physics,
Massachusetts Institute of Technology, Cambridge, Massachusetts 02139, USA}

\author{Furkan~\c{C}a\u{g}r{\i}~Top}
\affiliation{Research Laboratory of Electronics, MIT-Harvard Center for Ultracold Atoms, Department of Physics,
Massachusetts Institute of Technology, Cambridge, Massachusetts 02139, USA}

\author{Edward Su}
\affiliation{Research Laboratory of Electronics, MIT-Harvard Center for Ultracold Atoms, Department of Physics,
Massachusetts Institute of Technology, Cambridge, Massachusetts 02139, USA}

\author{Jeongwon Lee}
\affiliation{Research Laboratory of Electronics, MIT-Harvard Center for Ultracold Atoms, Department of Physics,
Massachusetts Institute of Technology, Cambridge, Massachusetts 02139, USA}

\author{Alan O. Jamison}
\affiliation{Research Laboratory of Electronics, MIT-Harvard Center for Ultracold Atoms, Department of Physics,
Massachusetts Institute of Technology, Cambridge, Massachusetts 02139, USA}

\author{Wolfgang Ketterle}
\affiliation{Research Laboratory of Electronics, MIT-Harvard Center for Ultracold Atoms, Department of Physics,
Massachusetts Institute of Technology, Cambridge, Massachusetts 02139, USA}

\date{\today}

\begin{abstract}
We propose and demonstrate a new approach for realizing spin-orbit coupling with ultracold atoms. We use orbital levels in a double-well potential as pseudospin states. Two-photon Raman transitions between left and right wells induce spin-orbit coupling. This scheme does not require near resonant light, features adjustable interactions by shaping the double-well potential, and does not depend on special properties of the atoms. A pseudospinor Bose-Einstein condensate spontaneously acquires an antiferromagnetic pseudospin texture which breaks the lattice symmetry similar to a supersolid.
\end{abstract}

\maketitle

Spin-orbit coupling is the mechanism for many intriguing phenomena, including $\mathbb{Z}_2$ topological insulators, the spin quantum Hall effect~\cite{Kane, Zhang}, Majorana fermions~\cite{Muller}, and spintronics devices~\cite{Spintronics}. Realizing controllable spin-orbit coupling with ultracold atoms should make it feasible to explore fundamental aspects of topology in physics and applications in quantum computing~\cite{Alicea2011}.

Spin-orbit coupling requires the atom's motion to be dependent on its spin state. Spin-orbit coupling without spin flips is possible for schemes which are diagonal in the spin component $\sigma_z$. Such spin-dependent vector potentials, which are sufficient for realizing quantum spin Hall physics and topological insulators, can be engineered using far-detuned laser beams to completely suppress spontaneous emission~\cite{Miyake2013,Aidelsburger2013a}.

However, spin flips (i.e., spin-orbit coupling terms involving $\sigma_x$ or $\sigma_y$ operators) are necessary for Rashba~\cite{Rashba} and Dresselhaus~\cite{Dress} spin-orbit coupling~\cite{Manchon2015}. Experiments with ultracold atoms couple pseudospin states using optical dipole transitions, which couple only to orbital angular momentum of the atom. Most realizations, including the first demonstration~\cite{Lin2011}, use hyperfine states of an alkali atom as pseudospins. In this case, the coupling of the two states occurs due to internal spin-orbit coupling in the excited state of the atom, which causes the fine-structure splitting between the $D_1$ and $D_2$ lines. The optimum detuning of the lasers is comparable to this splitting, leading to heating. Special atomic species with orbital angular momentum in the ground state can avoid this problem, as recently realized with dysprosium~\cite{Burdick2016}. Here we present a new method which can be applied to any atomic species, using an external orbital degree of freedom as pseudospin to avoid the need for near-resonant light.

\begin{figure}[h]
\includegraphics[width = 8.6cm]{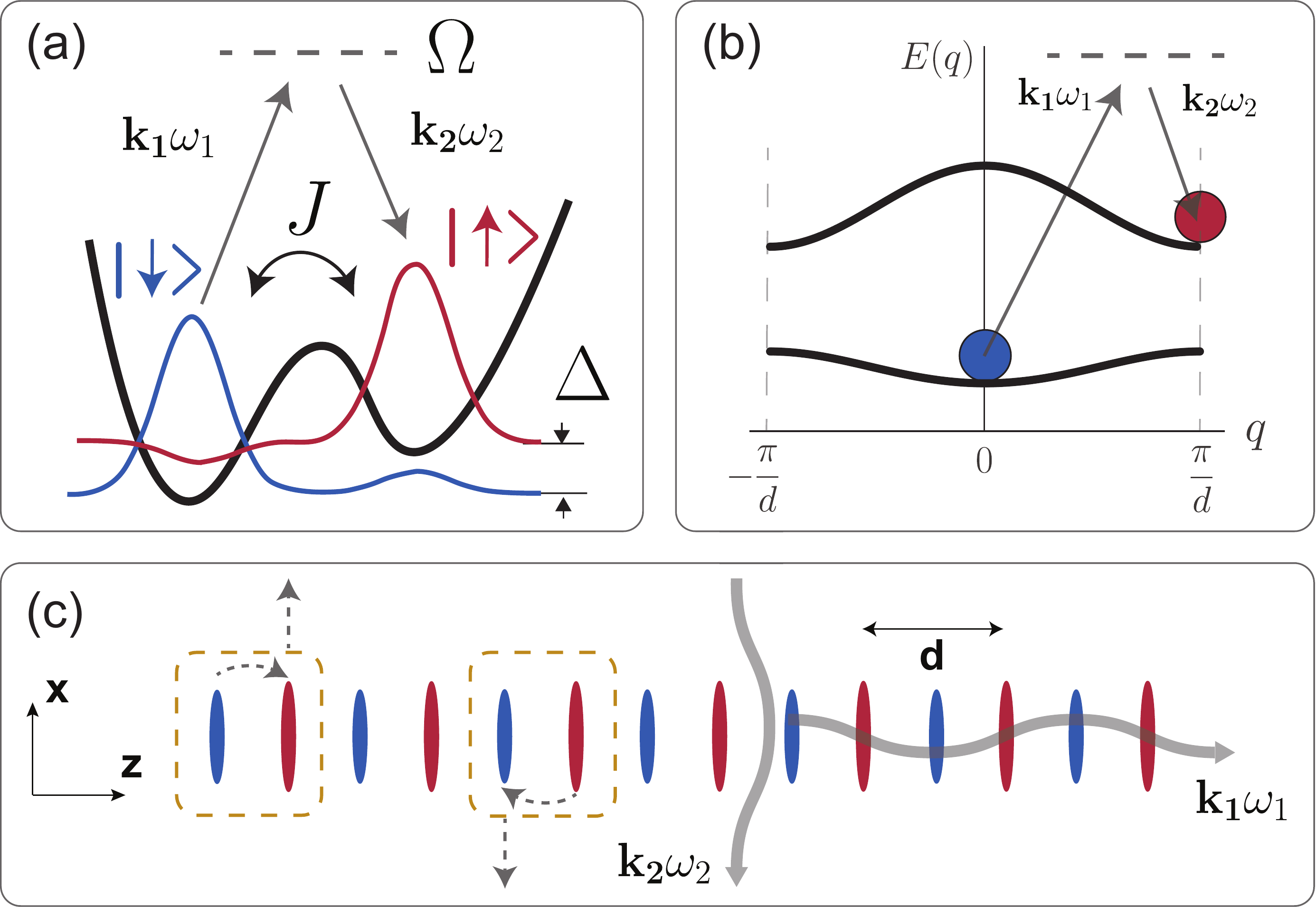}
\caption{(color)\label{fig:schematic}Realization of orbital pseudospins in a superlattice. (a)The unit cell of the superlattice is a double well with offset $\Delta$ and tunneling $J$. The two lowest eigenstates (pseudospin up and down) are coupled via a two-photon Raman process;(b)Raman process in the band structure of the superlattice. The ground state with quasimomentum $q=0$ is coupled to the edge of the Brillouin zone $q=\frac{\pi}{d}$ of the first excited band; (c)Top view of the superlattice with period $d=\lambda_{\rm IR}/2=532\;\rm nm$. Raman coupling is implemented by two $\lambda_{\rm IR}$ beams: one along the superlattice ($z$ direction), the other along the $x$-direction. SOC (curved arrows) transfers transverse recoil in the $x$-direction to the atoms (dashed arrows).}
\end{figure}

An external degree of freedom as pseudospin could be realized for a two-dimensional system by using the ground and first excited states of the confinement along the third dimension as pseudospin states. However, the excited state would rapidly relax due to elastic collisions, typically on a millisecond time scale~\cite{Spielman2006}. This is also the case for the recent implementation of SOC with hybrid $s-p$ Floquet bands in a one-dimensional optical lattice~\cite{Khamehchi2016}. To solve this issue, we choose an asymmetric double-well potential~(Fig.\;\ref{fig:schematic}).Pseudospins up and down are realized as the two lowest eigenstates of the double-well potential. For $J/\Delta \ll 1$, they can be expressed by the tight-binding states $\left|l\right>$ and $\left|r\right>$ localized in the left and right wells, respectively:  $\left|\downarrow\right>=\left|l\right>+\frac{J}{\Delta}\left|r\right>$ and $\left|\uparrow\right>=\left|r\right>-\frac{J}{\Delta}\left|l\right>$. The tunneling $J$ and offset $\Delta$ between the two wells are used to adjust the overlap---and therefore interactions and collisional relaxation rate---between the two pseudospin states. We couple the two states via a two-photon Raman transition with large detunings to achieve SOC with spin flips. (For convenience, we will refer to pseudospin as spin in this paper.) Recent work on two-leg ladders can be mapped to SOC between the two legs of the ladder~\cite{DHugel2014, MAtala2014}. Our scheme is qualitatively different from other realizations of orbital pseudospin since it realizes spin-orbit coupling in free space as compared to lattice models.

An intriguing prediction for spin-orbit coupled Bose-Einstein condensates (BECs) is the existence of a stripe phase~\cite{YLi2012,CWang2010,THo2011}, a spontaneous density modulation which realizes a supersolid~\cite{Boninsegni2012}. However, when the inter-spin ($g_{\uparrow\downarrow}$) and intra-spin ($g_{\uparrow\uparrow}$, $g_{\downarrow\downarrow}$) interaction strengths are the same, increased interaction energy of the density modulation drives spatial phase separation, eliminating the stripes. The system can be kept in the miscible phase when inter-spin interactions are weaker than intra-spin interactions
$g_{\uparrow\downarrow}^2<g_{\uparrow\uparrow}g_{\downarrow\downarrow}$~\cite{YLi2012}.
In our realization, $g_{\uparrow\downarrow}$ is proportional to the overlap squared of the wavefunctions on the two sides of the double-well. An analogous scheme can be realized with hyperfine pseudospins and spin-dependent lattices~\cite{Martone2014}, but requires near-resonant light. Our scheme does not depend on specific atomic properties and addresses three challenges to realizing the stripe phase: (1) Spin-orbit coupling without near resonant light, (2) miscible system with adjustable inter-spin interactions, (3) long lifetime against collisional relaxation.

Instead of one double-well system, we create a lattice of double wells using an optical superlattice~(Fig.\;\ref{fig:schematic}(c)). The advantages of working with a stack of coherently coupled double wells are twofold: increased signal to noise ratio and use of interference between the double wells to separately observe the two spin states. In the present work, the degree of freedom along the superlattice direction is purely an aid to observation~\cite{Note1}.

Our main result is the observation of the momentum structure of a BEC modified by a superlattice and spin-orbit coupling (SOC). We first describe the effects of the superlattice \textit{without} adding SOC. A one-dimensional superlattice of double wells was realized by combining lattices of $\lambda_{\rm IR}=1064\;\rm nm$ light and $\lambda_{\rm Gr}=532\;\rm nm$ light obtained by frequency doubling the  $\lambda_{\rm IR}=1064\;\rm nm$ light. The shape of the double-well unit cell is determined by the relative strength and spatial phase $\phi_{\rm SL}$  between the two lattices. The phase is controlled by a rotatable dispersive glass plate and an acousto-optical modulator for rapidly switching the IR lattice frequency.

\begin{figure}[h]
\includegraphics[width = 8.6cm]{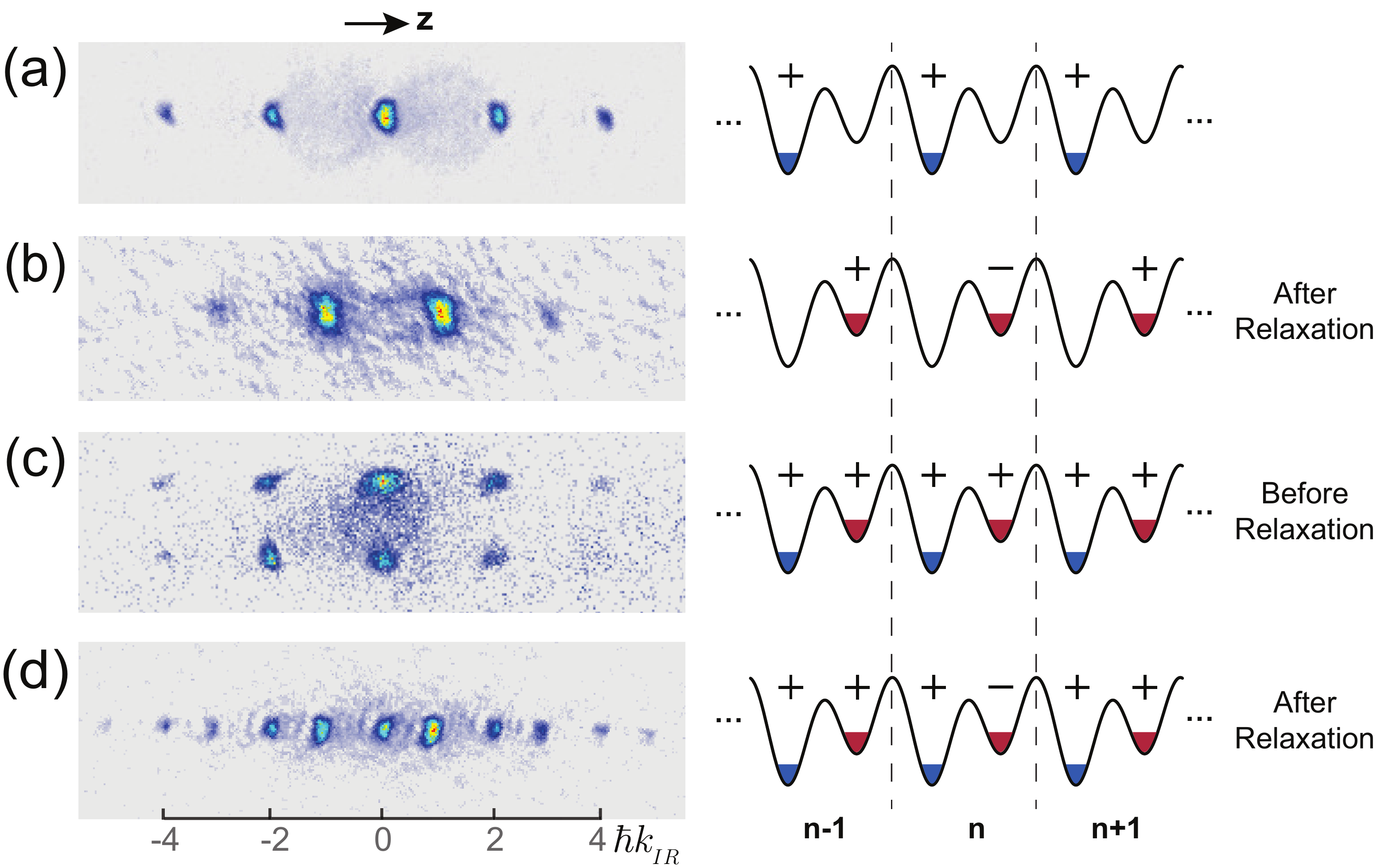}
\caption{(color)\label{fig:SternGerlach} Spontaneous formation of an antiferromagnetic spin texture. (a),(b)Time-of-Flight(TOF) pattern of atoms in the ground (first excited) band of the superlattice. After preparation of the $\ket{\uparrow}$ state with quasimomentum $q=0$, it relaxes to the bottom of the band at $q=\pi/d$. (c)An equal mixture of spin states is prepared by rapidly switching the superlattice parameters. The two spin states can be separated in TOF by a pseudospin Stern-Gerlach effect~\cite{Note2}.The figure shows that both spin states are in $q=0$ before the relaxation. (d)After relaxation, spinor BECs with states $\ket{\downarrow}$, $q=0$ and $\ket{\uparrow}$, $q=\pi/d$ are observed. The momentum pattern implies a periodic structure at $2d$, twice the lattice constant, indicating that an antiferromagnetic spin structure with a doubled unit cell has formed. The plus/minus signs indicate (one possible choice for) the phase of the BEC wavefunction. $n$ is the site index.}
\end{figure}

The experiment starts with a BEC of $\sim3\times10^{5}$ $^{23}$Na atoms in $\ket{F=1,m_{F}=-1}$ state in a crossed optical dipole trap. The superlattice is adiabatically ramped up within $250\;\rm ms$. For an offset $\Delta \gg J$, all the atoms equilibrate at the band minimum $q=0$ of the lowest superlattice band, putting $100\%$ of the population in the $\ket{\downarrow}$ state. The relative population of the two spin states can be controlled by first adjusting $\Delta$ for the loading stage to achieve a desired state population and then rapidly lifting one well up to the final offset~\cite{Note2}. The upper well corresponds to the first excited band which has its minimum energy at quasimomentum $q=\pi/d$ with $d = \lambda_{\rm IR}/2$ (\ref{fig:SternGerlach}(a),(b)). Since the lowest energy $\ket{\uparrow}$ and $\ket{\downarrow}$ states have different quasimomenta and experience different transverse confinement, they can be separately observed in ballistic expansion images without the band-mapping techniques~\cite{Note2}.

The $\pi/d$ quasimomenta difference also leads to an interesting spin texture for an equal population of $\ket{\uparrow}$ and $\ket{\downarrow}$ states. For this, atoms are prepared in both bands with $q=0$~(Fig.\;\ref{fig:SternGerlach}(c)), corresponding to a wavefunction periodicity of $532\;\rm nm$, i.e., the lattice constant. However, after relaxation, the periodicity has doubled to $1064\;\rm nm$, as indicated by the doubled number of momentum components in ballistic expansion images~(Fig.\;\ref{fig:SternGerlach}(d)). Specifically, the system was prepared in the symmetric state $\sum_{n}(\ket{\downarrow_{\rm n}}+\ket{\uparrow_{\rm n}})$ where $n$ denotes the lattice site, which is a ferromagnetic spin state in the $x-y$ plane. After relaxation into the state $\sum_{n}(\ket{\downarrow_{\rm n}}+(-1)^{n}e^{i\theta}e^{-i\Delta t}\ket{\uparrow_{\rm n}})$ an antiferromagnetic spin texture has developed which reduces the translational symmetry of the lattice. This system breaks both U(1) symmetry (the phase of the BEC) and the translational symmetry of the superlattice. In addition to the spin-density wave, it also has a density wave with the same period due to the interference of the $\ket{\uparrow}$ and $\ket{\downarrow}$ satellites. The position of the spin and density modulations is determined by the spontaneous phase $\theta$ and oscillate at frequency $\Delta$ ~\cite{Note2}. It is a simple system fulfilling one definition of supersolidity~\cite{Note3,Baumann2010,Chen2012}.

The small satellites allow spin-orbit coupling, but also lead to collisional decay of the $\left|\uparrow\right>$ state. We observed lifetimes on the order of $200\;\rm ms$ for both the $\ket{\uparrow}$ and equally mixed states at a density of $n\approx 2.5\times10^{14}\;\rm cm^{-3}$. The similar lifetimes for both states and its sensitivity to daily alignment indicate the lifetime being limited by technical noise and misalignment of the lattice rather than by collisions. Collisions would lead to a shorter lifetime for the mixed state by a factor of $4(J/\Delta)^2$. Adding Raman beams (with the parameters presented in Fig.\;\ref{fig:AB}) increases the loss rate by $\sim10/s$, probably caused by technical issues. While previous work with $^{87}$Rb reports a lifetime of seconds~\cite{Lin2011}, the Raman hyperfine spin flip scheme is not promising for lighter atoms because of substantially higher heating rates compared with $^{87}$Rb, which are 10$^3$(10$^5$) times higher for $^{23}$Na($^6$Li)~\cite{Luo2016}. Even without major improvements, the lifetimes achieved in our work are longer than any relevant dynamic time scale and should be sufficient for further studies, including observation of the stripe phase~\cite{Note4}.

Coupling between the two spin states is provided by two $\lambda_{\rm IR}$ beams: one along the superlattice direction $z$, the other orthogonal to it (along $x$). The frequency difference of the two beams is close to the offset in the double well, allowing near-resonant population transfer. The recoil $k_z$ along the lattice is necessary to couple the two orthogonal spin states in the double well, and was chosen to be $k_z=\pi/d$. The recoil kick $k_x$ in the transverse plane provides the coupling between the free-space motion in the transverse plane and the spin. It has opposite signs for the transition $\left|\downarrow\right>$ to $\left|\uparrow\right>$ and the reverse transition.

The Raman coupling can be described as a moving potential $V_{\rm Raman}=\Omega \cos(k_xx+k_zz-\delta\cdot t)$, characterized by a 2-photon Rabi frequency $\Omega$, a detuning of Raman beams $\delta$ and a wave vector $(k_x, 0, k_z)$. We characterize the states by their spin, quasimomentum $q$, and $x$ momentum $k_x$ (the $y$ momentum is always zero).

If the system is initially prepared in the state $\ket{\downarrow, q=0, k_x=0}$, the adiabatically ramped Raman beams will transfer it to a new eigenstate:
\begin{multline}
\label{equ:down}
    \ket{\Psi_1}=\ket{\downarrow,0,0}+K_1e^{-i\delta t}\ket{\uparrow,\pi/d, k_x}\\
    +M_1e^{-i\delta t}\ket{\downarrow, \pi/d, k_x}+M_1'e^{i\delta t}\ket{\downarrow, -\pi/d, -k_x}
\end{multline}
If prepared in $\ket{\uparrow, \pi/d, 0}$, the new state will be:
\begin{multline}
\label{equ:up}
    \ket{\Psi_2}=e^{-i\Delta t}\ket{\uparrow,\pi/d,0}+K_2e^{i(\delta-\Delta)t}\ket{\downarrow,0,-k_x}\\
    +M_2e^{i(\delta-\Delta)t}\ket{\uparrow,0,-k_x}+M_2'e^{-i(\delta+\Delta)t}\ket{\uparrow,0,k_x}
\end{multline}
The amplitudes obtained from 1$^{\rm st}$ order perturbation theory appear in Table I.
\begin{table}\label{tab:coe}
\caption{The amplitudes of the wavefunctions in eqs.\;(1)(2) obtained from 1$^{\rm st}$ order perturbation theory. (i = 1,2)}\label{table:amp}
\begin{ruledtabular}
\begin{tabular}{@{}cccc}
States&$M_i$&$M_i'$ &$K$\\
\hline
&&\\
$\ket{\Psi_1}$ & $-\frac{1}{2}\frac{\Omega}{E_r-\delta}$ & $-\frac{1}{2}\frac{\Omega}{E_r+\delta}$ & $-i\frac{e^{-i\frac{\pi}{4}}}{\sqrt{2}}\frac{J}{\Delta}\frac{\Omega}{E_r+\Delta-\delta}$\\
&&\\
$\ket{\Psi_2}$ & $+\frac{1}{2}\frac{\Omega}{E_r+\delta}$ & $-\frac{1}{2}\frac{\Omega}{E_r-\delta}$ & $+i\frac{e^{i\frac{\pi}{4}}}{\sqrt{2}}\frac{J}{\Delta}\frac{\Omega}{E_r-\Delta+\delta}$\\
&&\\
\end{tabular}
\end{ruledtabular}
\end{table}
The spin-orbit coupling is described by the second term in (1), (2). In addition, the Raman beams act as a co-moving lattice and (in the limit $\delta \gg E_r$) create a moving density modulation in the two spin states, described by the third and fourth terms. The spin-orbit coupling shows a resonant behavior for $\delta \approx \Delta$---the range of interest for SOC---where the moving density modulation is non-resonant. Both contributions are proportional to $\Omega/\Delta$. The off-resonant counter-rotating spin flip term is proportional to $\Delta^{-2}$ and has been neglected. For $\delta \gg E_r$, all off-resonant amplitudes $M_i$, $M_i'$ become $\approx\Omega/\delta$. For $\delta=\Delta$ and both spin states populated, the spin-orbit admixture of $\ket{\Psi_1}$ is expected to form a stationary interference pattern with $\ket{\Psi_2}$ along $x$ with wavevector $k_x$, and vice versa which constitutes the stripe phase of spin-orbit coupled BECs in the perturbative limit. (In general, the periodicity of the stripes depends on $\beta$ and the atoms' interactions~\cite{YLi2012}.)

The resonant Raman coupling leads to the standard spin-orbit Hamiltonian~\cite{Note2}:
\begin{equation}
\hat{H}_{SOC}=\frac{(\hat{p}+\alpha\hat{\sigma}_z)^2}{2m}+\beta\hat{\sigma}_x+\delta_0\hat{\sigma}_z,
\end{equation}
which can be considered as equal contributions of Rashba and Dresselhaus interactions. The parameters $\alpha = -k_x/2$, $\beta=(1/\sqrt{2})\Omega J/\Delta$, and $\delta_0 = (\delta-\Delta)/2$ are independently tunable in our experiment.
\begin{figure}[h]
\includegraphics[width = 8.6cm]{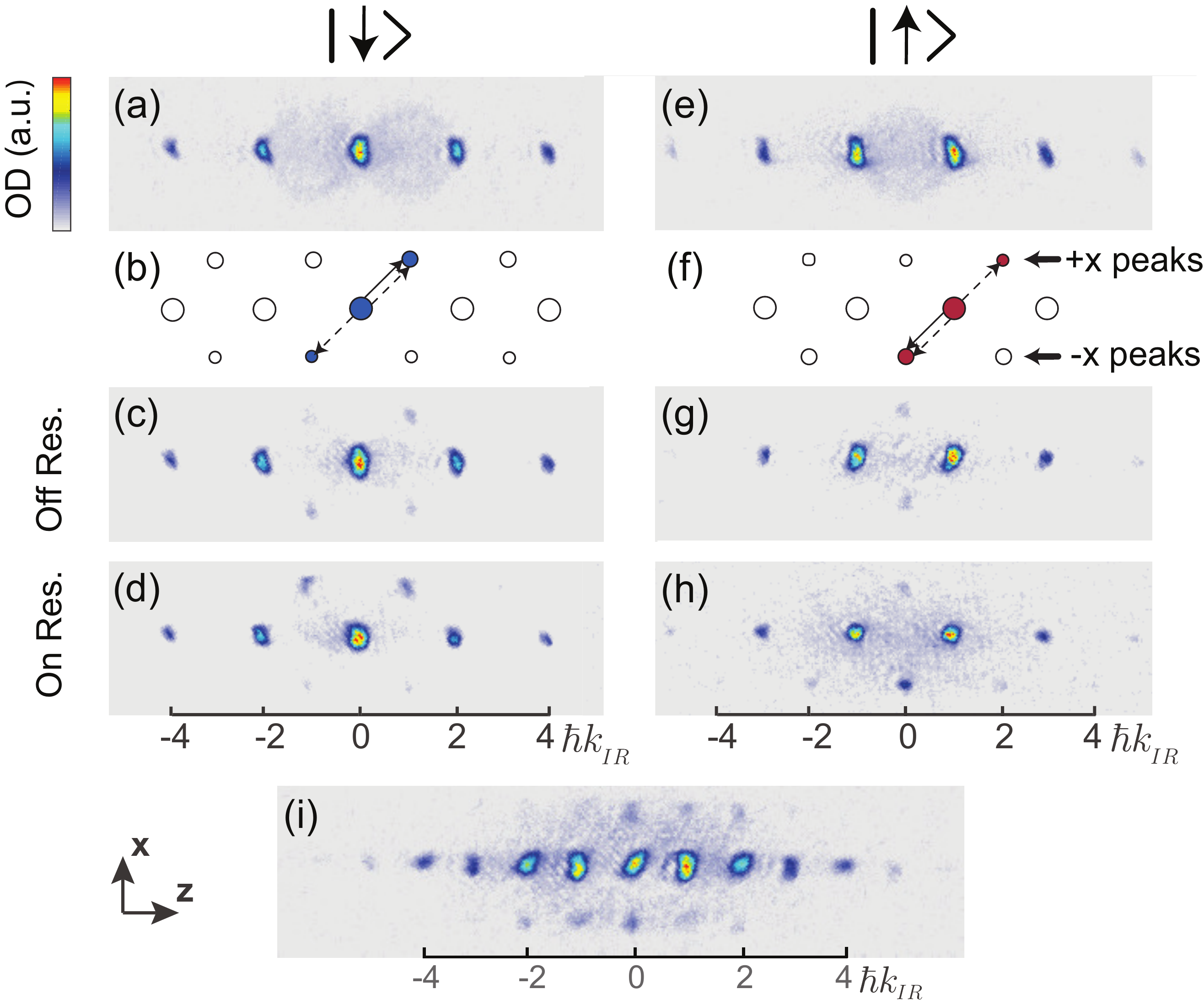}
\caption{(color)\label{fig:pattern} Characterization of spinor BECs through their momentum distributions. (a),(e) TOF images of $\ket{\downarrow}$ and $\ket{\uparrow}$ states, respectively. (b),(f) Schematics of the momentum peaks for $\ket{\downarrow}$ and $\ket{\uparrow}$ with Raman coupling. Both the SOC (solid arrows) and the density modulation (dashed arrows) are shown. The main peak (filled circle) is equal to the quasimomentum of the state. Extra peaks(open circles) appear due to the periodic potential. (c),(d),(g),(h)
Same as in (a) and (e), but now with Raman coupling at different detunings $\delta$. Momentum components created by Raman process are vertically shifted compared to (a) and (e) due to the transverse momentum kick. The momentum shift along the superlattice ($z$ direction) reflects the $\pi/d$ quasimomentum of the Raman lattice. The off-resonant density modulation creates momentum peaks which are symmetric along $+x$ and $-x$~(Figs. (c),(g)), whereas resonant spin-orbit coupling creates unidirectional momentum transfer resulting in asymmetry\;(Figs. (d),(h)). (i) Spin-orbit coupled BEC with equal population in spin up and spin down states}
\end{figure}

To characterize all the components of the wavefunctions above, the Raman coupling was adiabatically switched on by ramping up the intensity of the two Raman beams. The momentum space wavefunction was observed by suddenly switching off the lattice and trapping beams and measuring the resulting density distribution with absorption imaging after $10\;\rm ms$ of ballistic expansion~(Fig.\;\ref{fig:pattern}).

The momentum components created by the Raman beams are displaced in the $x$ by the recoil shift $\hbar k_{\rm IR}$. For off-resonant Raman beams, the pattern is symmetric for the $+x$ and $-x$ directions---signifying the moving density modulation (see (1)(2)). The resonant spin-orbit coupling is one-sided, with opposite transfer of $x$-momentum for the two spin states---as observed in Fig.\;\ref{fig:pattern}. We separate the momentum peaks due to the moving density modulation from SOC by evaluating the difference between the momentum peaks along the $+x$ and $-x$. Fig.\;\ref{fig:AB} shows the resonance feature of SOC when the Raman detuning was varied. The resonances for the two processes $\ket{\downarrow}\rightarrow \ket{\uparrow}$ and $\ket{\uparrow} \rightarrow \ket{\downarrow}$ should be separated by $2E_{r}\approx15.3\;\rm kHz$. The observed discrepancy is consistent with mean field interactions which reduce the separation by $\sim 2\mu \approx 5\;\rm kHz$, where $\mu$ is the single site chemical potential. The observed widths of the resonances are probably dominated by the inhomogeneity of $\Delta$ due to the Gaussian beam profile of the IR lattice laser~\cite{Note2}.
\begin{figure}[h]
\includegraphics[width = 8.6cm]{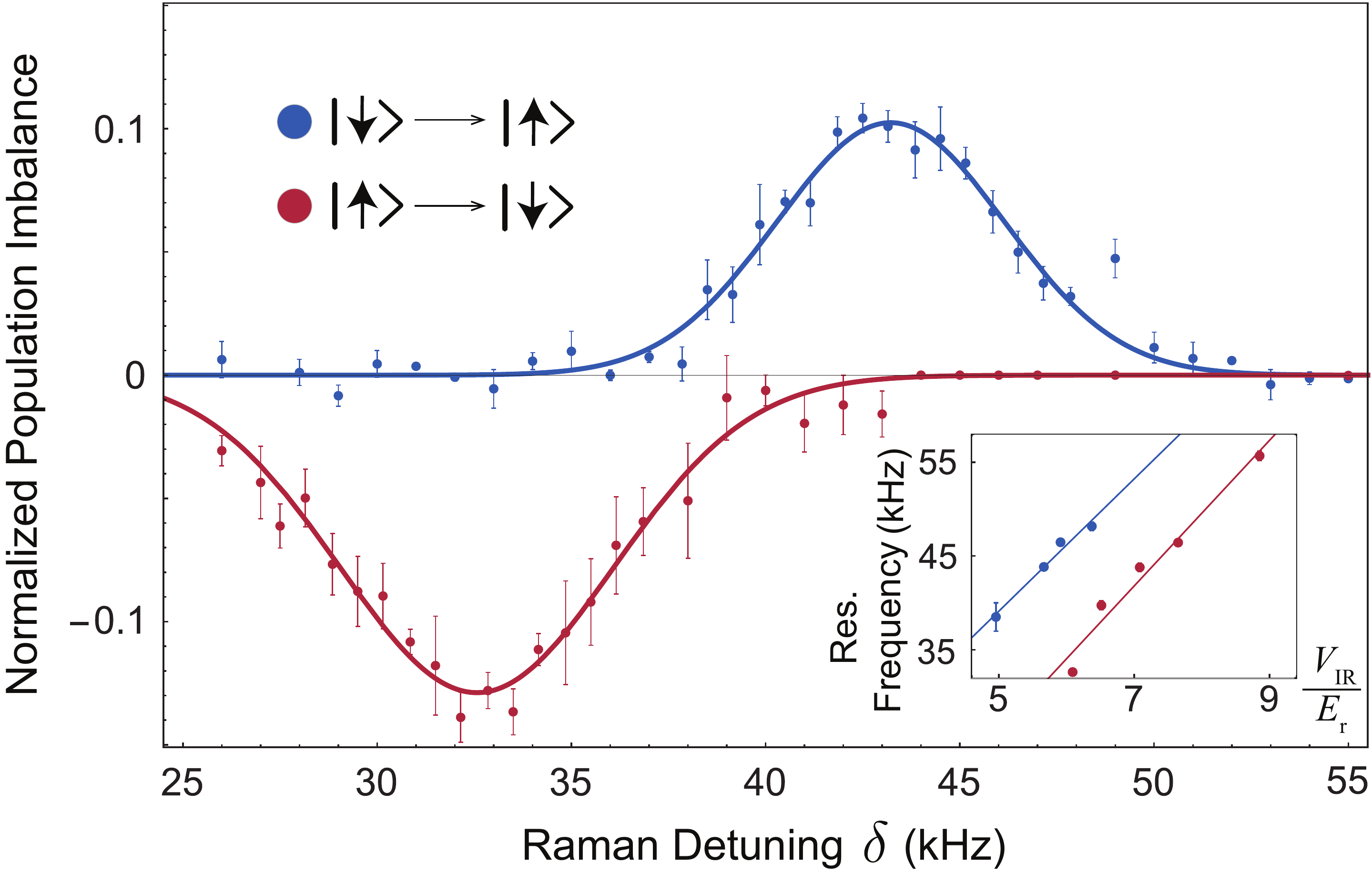}
\caption{(color)\label{fig:AB} Spin-orbit coupling resonances. Shown is the population imbalance between the ``$+x$" and ``$-x$" momentum peaks versus Raman detuning for $\ket{\downarrow} \rightarrow \ket{\uparrow}$(blue) and $\ket{\uparrow} \rightarrow \ket{\downarrow}$(red) processes. The two sets of data were measured for the same superlattice parameters $V_{\rm IR}= 7.5(2)E_{r}$, $V_{\rm Gr}=20(2)E_{r}$ and $\phi_{\rm SL} \approx 0.22(1)\pi$ which gives $\Delta \approx 37(1)\;\rm kHz$. The spin-orbit coupling strength $\beta$ was calculated to be $0.40(5) \;\rm kHz$. The solid lines are Gaussian fits to the resonances centered at $32.2(3)$kHz and $43.2(3)\;\rm kHz$. The Gaussian profile of the IR lattice inhomogeneously broadens the resonances. The error bars represent one $\sigma$ statistical uncertainty. Inset: Resonance center frequencies versus IR lattice depth $V_{\rm IR}$ for fixed $\phi_{\rm SL}$. The resonances are linear in $V_{\rm IR}$ with a constant split equal to twice the recoil energy. The slope of the linear fit reveals $\phi_{\rm SL}$. Error bars are the uncertainties of the fit.}
\end{figure}

Having established spin-orbit coupling at the single-particle level, the next step is to explore the phase diagram of spin-orbit coupled Bose-Einstein condensates with interactions \cite{YLi2012,Martone2014,THo2011}, particularly the stripe phase. The clear signature of the stripe phase is the stationary, periodic density modulation on the BEC mentioned above. The periodicity is tunable through the spin-orbit coupling strength and can be directly observed via Bragg scattering~\cite{Miyake2011}. In contrast to experiments carried out with $^{87}$Rb, which has similar inter- and intra-spin scattering lengths, our system has an adjustable inter-spin interaction $g_{\downarrow\uparrow}\approx (J/\Delta)^{2} g_{\downarrow\downarrow} = (J/\Delta)^{2} g_{\uparrow\uparrow}$. Small values of $g_{\downarrow\uparrow}/g_{\uparrow\uparrow}$ lead to a large window of $\beta$ for observing the stripe phase and enable higher contrast stripes~\cite{YLi2012}. Fig.\;\ref{fig:pattern}(i) shows the momentum distribution of an equal spin mixture with SOC. We observed $\sim40\; \rm ms$ lifetime for parameters presented in Fig.\;\ref{fig:AB}. After adding Bragg detection, the observation of the stripe phase is in reach. 

In conclusion, we proposed and demonstrated a new scheme for realizing spin-orbit coupling using superlattices. An asymmetric double-well potential provides attractive features for pseudospins, including long lifetimes, adjustable interactions, and easy detection. This scheme can be applied to a wide range of atoms including lithium and potassium, which suffer from strong heating when hyperfine pseudospins are coupled. On the other hand, by combining multiple hyperfine states with the orbital degree of the double well, our scheme can realize two-dimensional Rashba spin-orbit coupling~\cite{Su2016} and suggestions made for alkaline-earth atoms, for example synthetic non-abelian gauge potentials~\cite{Dalibard2011,Osterloh2005}, and Kondo lattice models~\cite{Gorshkov2010, Foss-Feig2010, Nakagawa2015}.

\begin{acknowledgments}
We acknowledge helpful discussions with S. Stringari, E. Mueller and N. Cooper. We thank A. Keshet for contributions to building the experiment and J. Amato-Grill, C. J. Kennedy and P. N. Jepsen for suggestions and critical reading of the manuscript. We acknowledge support from the NSF through the Center for Ultracold Atoms and by award 1506369, from ARO-MURI Non-equilibrium Many-body Dynamics (grant W911NF-14-1-0003) and from AFOSR-MURI Quantum Phases of Matter (grant FA9550-14-1-0035).
\end{acknowledgments}

\newpage
\onecolumngrid
\section{Superlattice Hamiltonian with Raman coupling}
The Hamiltonian for a one-dimensional superlattice, created by standing waves of infrared and green light with relative phase $\phi_{SL}$ is

$$H_{lattice}=\frac{\hat{p}_z^2}{2m}+\frac{\hat{p}_{\perp}^2}{2m}+V_{Gr}\sin^2(k_{Gr}z)+V_{IR}\sin^2(k_{IR}z+\phi_{SL})$$

It can be rewritten as
$$H_{lattice}=\frac{\hat{p}_{\perp}^2}{2m}+\frac{1}{2}\Delta_0\sum\limits_{n}(\left|r_n\right>\left<r_n\right|-\left|l_n\right>\left<l_n\right|)-J\sum\limits_{n}(\left|l_n\right>\left<r_n\right|+h.c.)-\sum\limits_{n}\sum\limits_{\substack{t=l,r\\t'=l,r}}(J'_{tt'}\ket{t_n}\bra{t'_{n+1}}+h.c.),$$
where $\ket{l_n(r_n)}$ is a wavefunction localized in the left(right) well of the $n^{th}$ unit cell, $\Delta_0$ is the energy separation between the right and the left wells. $\hbar$ is taken to be 1. Tunneling between neighboring unit cells is important for maintaining of coherence in the superlattice, but not relevant for the physics of spin-orbit coupling. Thus, tunneling terms with $J'_{tt'}$ in the Hamiltonian can be neglected.

The complete Hamiltonian of the system is
$$H=H_{lattice}+V_{Raman},$$ where $V_{Raman}=\Omega\cos(k_zz+k_xx-\delta t)$ is a moving lattice potential. For the purpose of our experiment, the initial phase of the Raman potential is not relevant and taken to be zero.

We prefer to use eigenstates of a double well for the description. To first order in small parameter $\frac{J}{\Delta_0}\ll1$ they can be written as $$\left|\downarrow_n\right>=\left|l_n\right>+\frac{J}{\Delta_0}\left|r_n \right>, \;\;\; \left|\uparrow_n\right>=\left|r_n\right>-\frac{J}{\Delta_0}\left|l_n \right>$$

We expand the Hamiltonian in the new basis with $\Delta=\Delta_0+2\frac{J^2}{\Delta_0}\approx \Delta_0$:
$$H=\frac{\hat{p}_{\perp}^2}{2m}+\frac{1}{2}\Delta\sum\limits_{n}(\left|\uparrow_n\right>\left<\uparrow_n\right|-\left|\downarrow_n\right>\left<\downarrow_n\right|)+\sum\limits_{p_{\perp}, p_{\perp}^{'}}\ket{p_{\perp}}\Big(\sum\limits_{n}\sum\limits_{\substack{i=\downarrow,\uparrow\\i'=\downarrow,\uparrow}}\left|i_n\right>\left<i_n\right|\bra{p_{\perp}}\Omega\cos(k_zz+k_xx-\delta\cdot t)\ket{p'_{\perp}}\left|i_n'\right>\left<i_n'\right|\Big)\bra{p'_{\perp}}$$
In our experiment $k_z\approx k_x\approx k_{IR}=\frac{\pi}{d}$, where $d$ is a period of the superlattice. In order to estimate the effect of the Raman potential with arbitrary phase we need to know the overlap integrals for $\cos(k_{IR}\cdot(z-z_n))$ and $\sin(k_{IR}\cdot(z-z_n))$. To first order in $\frac{J}{\Delta}$:

$$\left<\downarrow_n\right|\cos(k_{IR}\cdot(z-z_n))\left|\uparrow_n\right>\approx \left<l_n\right|-\frac{J}{\Delta}\left|l_n\right>=-\frac{J}{\Delta},$$

$$\left<\downarrow_n\right|\cos(k_{IR}\cdot(z-z_n))\left|\downarrow_n\right>\approx 1, \left<\uparrow_n\right|\cos(k_{IR}\cdot(z-z_n))\left|\uparrow_n\right>\approx 0,$$

$$\left<\downarrow_n\right|\sin(k_{IR}\cdot(z-z_n))\left|\uparrow_n\right>\approx \left<r_n\right|\frac{J}{\Delta}\left|r_n\right>=\frac{J}{\Delta},$$

$$\left<\downarrow_n\right|\sin(k_{IR}\cdot(z-z_n))\left|\downarrow_n\right>\approx 0, \left<\uparrow_n\right|\sin(k_{IR}\cdot(z-z_n))\left|\uparrow_n\right>\approx 1,$$
where $z_n=nd$ is a coordinate of the left well in the $n^{th}$ unit cell.

Thus, the Raman potential can be expanded in the basis of double-well eigenstates:

\begin{multline*}
\sum\limits_{\substack{i=a,b\\i'=a,b}}\left|i_n\right>\left<i_n\right|\Omega\cos(k_z(z-z_n)+k_zz_n+k_xx-\delta\cdot t)\left|i_n'\right>\left<i_n'\right|=\\
=\Omega\cos\phi_n\{-\frac{J}{\Delta}\big(\left|\downarrow_n\right>\left<\uparrow_n\right|+\left|\uparrow_n\right>\left<\downarrow_n\right|\big)+\left|\downarrow_n\right>\left<\downarrow_n\right| \}+\\
-\Omega\sin\phi_n\{\left|\uparrow_n\right>\left<\uparrow_n\right|+\frac{J}{\Delta}(\left|\downarrow_n\right>\left<\uparrow_n\right|+\left|\uparrow_n\right>\left<\downarrow_n\right|)\},
\end{multline*}

where  $\phi_n=\pi n+k_xx-\delta t$. Later, we will calculate how $x$ as an operator acts on the momentum states $\ket{p_{\perp}}$.

\begin{equation}
\hat{V}_{Raman}=\sum\limits_{n}\Omega(-1)^n\cos(k_xx-\delta t) \left|\downarrow_n\right>\left<\downarrow_n\right|-\Omega(-1)^n\sin(k_xx-\delta t)\left|\uparrow_n\right>\left<\uparrow_n\right|+\\
\end{equation}
\begin{equation}
-\sqrt{2}\Omega\frac{J}{\Delta}(-1)^n\cos(k_xx-\delta t -\frac{\pi}{4})\{\left|\downarrow_n\right>\left<\uparrow_n\right|+\left|\uparrow_n\right>\left<\downarrow_n\right|\}
\end{equation}

The factor $(-1)^n$ represents the phase of the Raman beams, which have a wavelength two times the length of the unit cell.
In our experiment the atomic sample is prepared in the zero-momentum state. When the Raman perturbation is applied the atoms experience a kick in the $x$-direction. In the $y$-direction atoms remain unperturbed, i.e. $\hat{p}_y=0$. Since the confinement along $x$ is weak, we can use the basis $\left|\uparrow(\downarrow), k\right>=\left|\uparrow(\downarrow)\right>\otimes e^{ikx}$. The Raman interaction gives rise to intra-band coupling terms (1), and to the spin-orbit coupling term (2).

 The system in our experiment initially prepared in the lower wells, which corresponds to the $q=0$ of the lowest band of the superlattice:

$$\ket{\psi_{q=0}^{(\downarrow)}}=\sum\limits_{n=1}^{N}\frac{1}{\sqrt{N}}\left|\downarrow_n\right>$$

$N$ is the number of unit cells in the lattice. When all the atoms are confined in the upper wells, the lowest state is the $q=\frac{\pi}{d}$ state of the first excited band, due to the inverted dispersion relation:
$$\ket{\psi_{q=\pi/d}^{(\uparrow)}}=\sum\limits_{n=1}^{N}\frac{1}{\sqrt{N}} e^{i\frac{\pi}{d}(z_n+\frac{d}{2})}\left|\uparrow_n\right>$$

With $z_n=nd$, the state becomes:
$$\ket{\psi_{q=\pi/d}^{(\uparrow)}}=\sum\limits_{n=1}^{N}\frac{1}{\sqrt{N}} i (-1)^n\left|\uparrow_n\right>$$
For the calculation of matrix elements we assume that overlap is nonzero only for $n=n'$.

Intra-band coupling terms:

\begin{multline}
\left<\psi_{q=\pi/d}^{(\downarrow)}\right|\hat{V}_{Raman}\left|\psi_{q=0}^{(\downarrow)}\right>=\sum_{n,n'}\frac{1}{N}(-1)^n\left<\downarrow_{n'}\right|\Omega\cos(k_zz+k_xx-\delta t)\ket{\downarrow_n}\\
=\Omega\sum_n\frac{1}{N}\cos(\phi_n)(-1)^n=\Omega\cos(k_xx-\delta t)
\end{multline}

\begin{multline}
\left<\psi_{q=0}^{(\uparrow)}\right|\hat{V}_{Raman}\left|\psi_{q=\pi/d}^{(\uparrow)}\right>=\sum_{n,n'}\frac{1}{N}\left<\uparrow_{n'}\right|\Omega\cos(k_zz+k_xx-\delta t)i(-1)^n\ket{\uparrow_n}\\
=\Omega\sum_n\frac{1}{N}i(-1)^n(-\sin(\phi_n))=-i\Omega\sin(k_xx-\delta t)
\end{multline}

Spin-orbit coupling matrix element:

\begin{multline*}
\left<\psi_{q=\pi/d}^{(\uparrow)}\right|\hat{V}_{Raman}\left|\psi_{q=0}^{(\downarrow)}\right>=\sum_{n,n'}\frac{1}{N}(-i)(-1)^{n'}\left<\uparrow_{n'}\right|\Omega\cos(k_zz+k_xx-\delta t)|\left|\downarrow_n\right>\\
=-i\Omega\sum_n\frac{1}{N}(-1)^n\left<\uparrow_{n}\right|\cos(k_z(z-z_n))\cos(\phi_n)-\sin(k_z(z-z_n))\sin(\phi_n)\left|\downarrow_n\right>,
\end{multline*}

\begin{multline*}
\left<\psi_{q=\pi/d}^{(\uparrow)}\right|\hat{V}_{Raman}\left|\psi_{q=0}^{(\downarrow)}\right>=i\frac{J}{\Delta}\Omega\sum_n(-1)^n\frac{1}{N}\big(\cos(\phi_n)+\sin(\phi_n)\big)=i\frac{J}{\Delta}\Omega(\cos(k_xx-\delta t)+\sin(k_xx-\delta t))
\end{multline*}

Intra-band coupling matrix elements (3) and (4) provide recoil kick in $x-$direction with recoil energy $E_r=\frac{k_x^2}{2m}$ and along the superlattice, changing the quasimomentum by half a reciprocal vector.
If the system is initially at $\ket{\psi_{q=0}^{(\downarrow)}, 0}$, the new adiabatically connected eigenstate in first order perturbation theory is:
\begin{multline*}
\ket{\Psi_1}=\ket{\psi_{q=0}^{(\downarrow)}, 0}-\frac{1}{2}\frac{\Omega}{E_r-\delta}e^{-i\delta t}\ket{\psi_{q=\pi/d}^{(\downarrow)}, k_x}-\frac{1}{2}\frac{\Omega}{E_r+\delta}e^{i\delta t}\ket{\psi_{q=\pi/d}^{(\downarrow)},-k_x}+\\
-i\frac{e^{-i\frac{\pi}{4}}}{\sqrt{2}}\frac{J\Omega/\Delta}{E_r+\Delta-\delta}e^{-i\delta t}\ket{\psi_{q=\pi/d}^{(\uparrow)},k_x}-i\frac{e^{i\frac{\pi}{4}}}{\sqrt{2}}\frac{J\Omega/\Delta}{E_r+\Delta+\delta}e^{i\delta t}\ket{\psi_{q=\pi/d}^{(\uparrow)},-k_x}
\end{multline*}
If the system is prepared in $\ket{\psi_{q=\pi/d}^{(\uparrow)}, 0}$:
\begin{multline*}
    \ket{\Psi_2}=e^{-i\Delta t}\ket{\psi_{q=\pi/d}^{(\uparrow)}, 0}-\frac{1}{2}\frac{\Omega}{E_r-\delta} e^{-i(\delta+\Delta)t} \ket{\psi_{q=0}^{(\uparrow)}, k_x}+\frac{1}{2}\frac{\Omega}{E_r+\delta}e^{i(\delta-\Delta)t}\ket{\psi_{q=0}^{(\uparrow)},-k_x}+\\
+i\frac{e^{-i\frac{\pi}{4}}}{\sqrt{2}}\frac{\Omega J/\Delta}{E_r-\Delta-\delta}e^{-i(\delta+\Delta)t}\ket{\psi_{q=0}^{(\downarrow)}, k_x}+i\frac{e^{i\frac{\pi}{4}}}{\sqrt{2}}\frac{\Omega J/\Delta}{E_r-\Delta+\delta}e^{i(\delta-\Delta)t}\ket{\psi_{q=0}^{(\downarrow)}, -k_x}
\end{multline*}

For $\delta$ close to $\Delta$ intra-band coupling is off-resonant and both co- and counter-rotating terms contribute at comparable strengths, whereas for spin-orbit coupling the co-rotating term is resonant and, therefore, much stronger than the counter-rotating term.

\section{Spin-orbit-coupling Hamiltonian}
Keeping only the near-resonant spin-orbit coupling term, the Hamiltonian describing the system is
\begin{equation*}
H_{SOC}=
\begin{pmatrix}
\frac{\hat{p}_x^2}{2m}-\frac{\Delta}{2} & \frac{-ie^{i\pi/4}}{\sqrt{2}}\frac{J}{\Delta}\Omega e^{-i(k_xx-\delta t)}\\
\frac{ie^{-i\pi/4}}{\sqrt{2}}\frac{J}{\Delta}\Omega e^{i(k_xx-\delta t)} & \frac{\hat{p}_x^2}{2m}+\frac{\Delta}{2}
\end{pmatrix}
\end{equation*}
After a unitary transformation with a position-dependent rotation,  $\hat{U}=e^{(-ik_xx+i\delta t-i\frac{\pi}{4})\sigma_z/2}$, the Hamiltonian turns into $H_{SOC}^{'}=U^{\dagger}H_{SOC}U-i U^{\dagger}\frac{\partial U}{\partial t}$.

$$H_{SOC}^{'}=\frac{(\hat{p}_x+\alpha\sigma_z)^2}{2m}+\beta\sigma_x+\delta_0\sigma_z,$$
where $\alpha=-\frac{1}{2}k_x$, $\beta=\frac{J\Omega}{\sqrt{2}\Delta}$ and $\delta_0=\frac{1}{2}(\delta-\Delta)$. For the total Hamiltonian we have to add the intra-band coupling, which causes a density modulation.

\section{Antiferromagnetic spin texture}
A spin-$1/2$ aligned along the $\cos\phi\hat{x}+\sin\phi\hat{y}$ direction has the wavefunction $\ket{\downarrow}+e^{i\phi}\ket{\uparrow}$.
In our experiment atoms can be prepared in left and right sites of the double wells with equal population. The wave function is proportional to
$$\ket{\psi}=\sum\limits_{n}\ket{\downarrow_n}+(-1)^ne^{i\theta}e^{-i\Delta t}\ket{\uparrow_n},$$
which corresponds to spin-states aligned in the x-y plane with opposite direction on neighboring sites, showing x-y antiferromagnetic ordering. The expectation values of spin evolve as $\left<\sigma_x\right>=(-1)^n\cos(\theta-\Delta t)$, $\left<\sigma_y\right>=(-1)^n\sin(\theta-\Delta t)$ and $\left<\sigma_z\right>=0$.

Tunneling between wells within a unit cell causes a density modulation. Local populations in the $n^{\rm th}$ cell acquire a density imbalance:
$$|\left<{l_n}|\psi\right>|^2\sim\big(1-(-1)^n\frac{J}{\Delta}\cos(\theta-\Delta t)\big),$$
$$|\left<{r_n}|\psi\right>|^2\sim\big(1+(-1)^n\frac{J}{\Delta}\cos(\theta-\Delta t)\big)$$

Translating by one superlattice period flips the sign of the imbalance. This shows that $r_n$ and $l_{n+1}$ wells, and $r_{n-1}$ and $l_n$ wells have the same change in density, but oscillate out of phase. Therefore, this density wave has twice the wavelength of the superlattice and is shifted spatially by half a unit cell.

\section{Superlattice Calibration}
The superlattice potential $V(x) = V_{\rm IR}\sin^{2}(k_{\rm IR}z+\phi_{\rm SL})+V_{\rm Gr}\sin^{2}(k_{\rm Gr}z)$ was produced by overlapping two one-dimensional lattices with spacing $\lambda_{\rm IR}/2 = 532\;\rm nm$(long) and $\lambda_{\rm Gr}/2=266\;\rm nm$(short).  The $532\;\rm nm$ light was generated by frequency doubling a high power $\lambda = 1064\;\rm nm$ laser seeded by the same seed laser as the laser for the long lattice. This eliminated the need for any phase or frequency locking of two independent lasers. The long and short lattice also shared the same retroreflective mirror to minimize noise and drifts in the relative phase $\phi_{\rm SL}$.
\begin{figure}[h]
\includegraphics[width = 8.6cm]{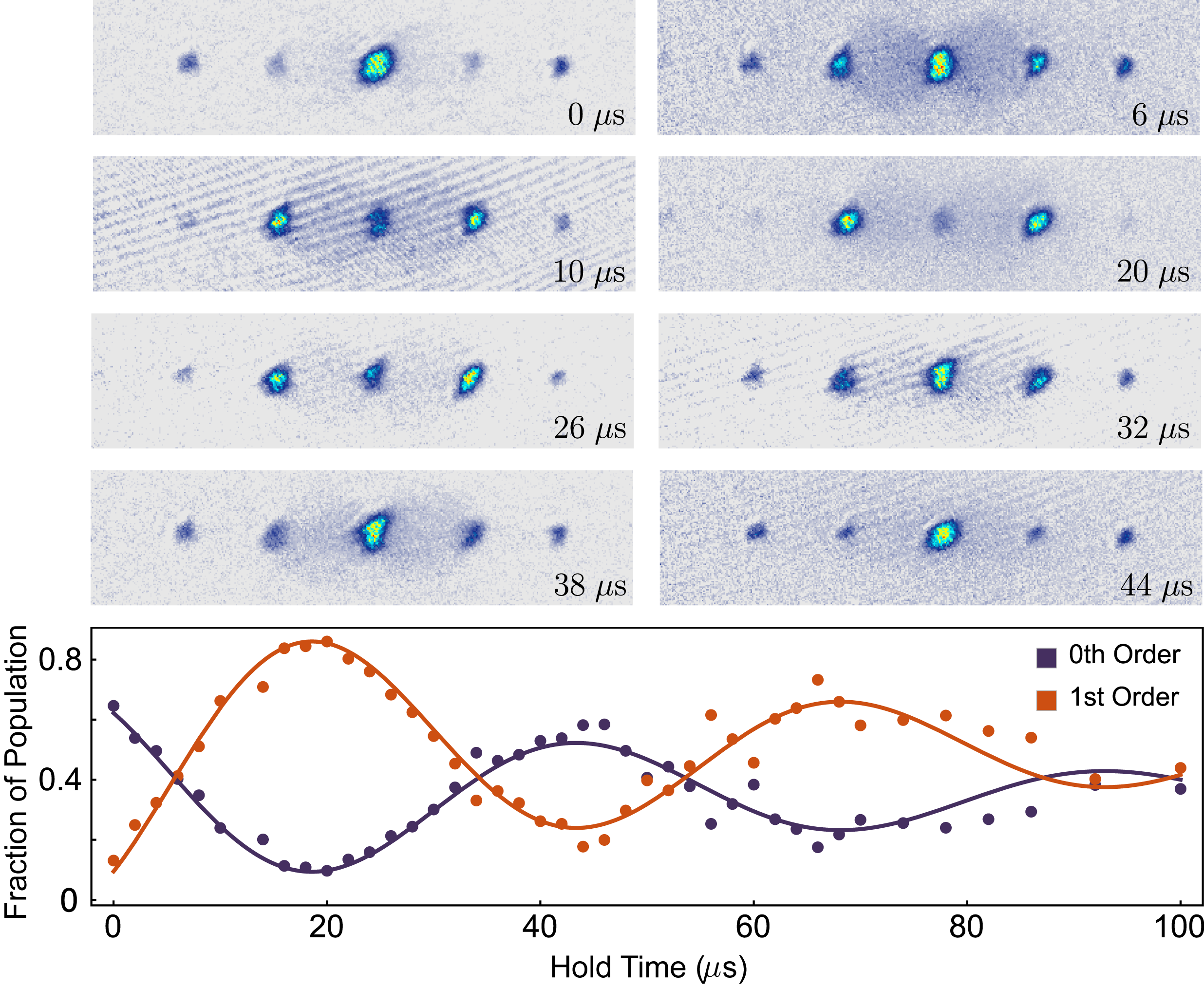}
\caption{~(color)\label{fig:pattern}Calibration of the superlattice offset $\Delta$. The relative phase between atoms in the lower and upper wells accumulated linearly with time and results in periodic changes in the time-of-flight interference pattern. The period of the oscillation is equal to $\Delta$ which is $\sim 23\;\rm kHz$ for the data shown.}
\end{figure}

 The relative phase $\phi_{\rm SL}$ was controlled by introducing different phase shifts for the $\lambda = 1064\;\rm nm$ and $532\;\rm nm$ lattice lights from a rotatable dispersive glass plate and by switching the frequency of the $1064\;\rm nm$ light. The glass plate allowed a wide tuning range for $\phi_{\rm SL}$. This design minimizes the optical path length, reducing the sensitivity to atmospheric pressure changes. The frequency switching allowed rapid but small phase shifts. For our geometry, a $70 \;\rm MHz$ shift for the $1064\;\rm nm$ light corresponded to a $\pi/4$ change in $\phi_{\rm SL}$

The offset $\Delta$ was directly calibrated by observing a beat note within a single double-well. Interference patterns for atom sitting in the lower and upper well overlap and therefore interfere with each other before they relax to orthogonal quasi-momentum states(Main Article Fig.3). The interference pattern evolves periodically with frequency $\omega = \Delta$(Fig.\ref{fig:pattern}). Condensates with equal population of atoms in the left and right well were prepared by pre-setting the glass plate to the desired value and then rapidly ramping up the IR lattice to the final offset.

\section{Pseudospin Stern-Gerlach Effect}
\begin{figure}[h]
\includegraphics[width = 8.6cm]{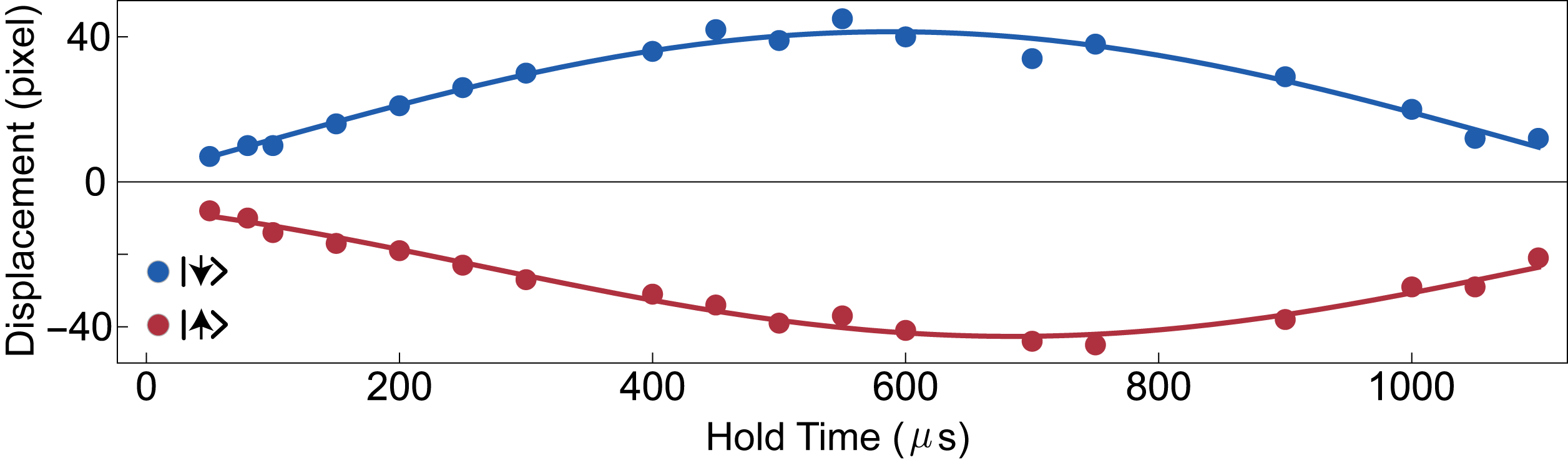}
\caption{~(color)\label{fig:slosh}Stern-Gerlach separation of the spin states. Atoms are loaded into the green lattice , then the IR lattice is suddenly ramped up. After different hold times, the vertical displacements of atoms in different spin states is measured with 10ms ballistic expansion. $y=0$ corresponds to the position of atoms without the sudden ramp. The spin states are displaced due to of transverse oscillations in the superlattice. The oscillating frequencies are equal to the transverse trapping frequencies which the spins experience in the superlattice. In the data shown, atoms in $\ket{\uparrow}$ and $\ket{\downarrow}$ oscillates with frequencies of $356\;\rm Hz$ and $417\;\rm Hz$ correspondingly.}
\end{figure}

Atoms in different pseudospin states can be spatially separated by a pseudospin Stern-Gerlach effect. The two spin states experience different transverse confinement from the $1064\;\rm nm$ lattice. When this lattice beam is displaced from the green lattice, the atoms in the two spin states experience different momentum kicks in the $x - y$ plane when the IR lattice is suddenly increased. This leads to transverse oscillations of the two spin states relative to each other and can be used to separate them in ballistic expansion(Fig.\ref{fig:slosh}). The frequency of the oscillation is equal to the corresponding transverse trapping frequency of each spin state.

\section{Resonance Line Broadening}

The spin-orbit coupling shows resonant behavior as a function of the frequency difference between the Raman beams. We attribute the broadening of the resonant line shape to 1) Doppler broadening due to the finite size of the trapped condensate 2) inhomogeneous mean-field shift due to inhomogeneous condensate density 3) inhomogeneous shift of the superlattice offset $\Delta$ due to the Gaussian beam profile.

First, the finite size of the condensate implies a distribution of momenta along the transverse direction which Doppler broadens the resonance. A momentum $\bf{p}$ along $x$ direction will shift the resonance by $\bf{k}_{\rm IR}\cdot \bf{p}/ m$ considering the geometry of our setup. Therefore a condensate size $x_0$ results in a broadening of $\sim k_{\rm IR}\hbar/mx_{0}$. Assuming a Thomas-Fermi distribution in the transverse direction, we obtain an rms width of\cite{Stenger1999}
$$
\delta \nu_{D} = \sqrt{\frac{21}{8}}\frac{k_{\rm IR}}{2\pi}\frac{\hbar}{mx_0}
$$
where $x_{0} = \sqrt{2\mu/m(2\pi\nu_{\rm x})^{2}}$. This gives a broadening of $\sim 400\;\rm Hz$.

The resonance is also broadened by the inhomogeneous density distribution of the condensate as
$$
\delta \nu_M = \sqrt{\frac{8}{147}}\frac{\mu}{h}
$$
with $\mu$ being the chemical potential. This effect broadens the resonance by $\sim 600\;\rm Hz$.

The Gaussian beam profile of the long lattice implies an inhomogeneous offset $\Delta$ within the sample, therefore broadening the transition. The resulted shift is estimated to be

$$
\delta\nu_I \approx 2\Delta (\frac{x_{0}}{\sigma})^{2}
$$
where $\sigma$ is the Gaussian beam waist parameter. This effect broadens the resonance line by $\sim 1.10\;\rm kHz$. However, the broadening can be much larger for small displacement between the green and IR lattice.

The three widths add up quadratically to a value of $1.30\;\rm kHz$.

The linewidth of the frequency difference of the Raman beams was negligible --- it was monitored through a beat note and was $\sim 50$ Hz without actively phase-locking the two.
We conclude that observed resonance width of $2\;\rm kHz$ is most likely dominated by the spread in $\Delta$ due to slight misalignment of lattice beams.

\end{document}